\documentclass[11pt]{article}

\usepackage[utf8]{inputenc}

\usepackage{geometry}
\usepackage{hyperref}
\usepackage{amsmath}
\usepackage{graphicx}
\usepackage{subcaption}

\usepackage[square,numbers]{natbib}
\usepackage{booktabs}
\usepackage{array}
\usepackage{paralist}
\usepackage{verbatim}

\usepackage{fancyhdr}
\usepackage{sectsty}
\allsectionsfont{\sffamily\mdseries\upshape}

\usepackage[nottoc,notlof,notlot]{tocbibind}
\usepackage[titles,subfigure]{tocloft}

\newcommand{\be}{\begin{equation}}
\newcommand{\ee}{\end{equation}}

\newcommand{\bs}{\boldsymbol}
 
\title{Optimal portfolio allocation with uncertain covariance matrix}

\author{Maxime Markov\thanks{Corresponding Author: \href{markov@theory.polytechnique.fr}{markov@theory.polytechnique.fr}}  \,\, and Vladimir Markov}

\date{\today}

\begin{document}
\maketitle

\begin{abstract}

In this paper, we explore the portfolio allocation problem involving an uncertain covariance matrix. We calculate the expected value of the Constant Absolute Risk Aversion (CARA) utility function, marginalized over a distribution of covariance matrices. We show that marginalization introduces a logarithmic dependence on risk, as opposed to the linear dependence assumed in the mean-variance approach. Additionally, it leads to a decrease in the allocation level for higher uncertainties. Our proposed method extends the mean-variance approach by considering the uncertainty associated with future covariance matrices and expected returns, which is important for practical applications. 
   
\end{abstract}

\section{Introduction}
Portfolio allocation is a classical problem in finance. The two dominant methodologies for deriving portfolio weights are either by maximizing the expected utility (EU)~\cite{Sharpe:2007} or by focusing on the mean-variance (MV) of the portfolio in relation to those weights~\cite{Markowitz:1952}. While the MV approach is favored for its conceptual simplicity and analytical tractability, the utility function method offers a more comprehensive perspective. When using a Constant Absolute Risk Aversion (CARA) utility function with Gaussian returns, both methods yield identical results. However, because real-world returns are not strictly Gaussian, the utility function method, which can accommodate arbitrary return distributions, is often preferred.

All modern portfolio optimization models use expectations of future returns and variances estimated from the past data. These estimates are imperfect and can be very far from reality during market regime changes. Comprehensive modeling and optimization should explicitly incorporate the fact that future returns and variances are unknown, utilize predictive models of future returns and variances, and study the implications of these uncertainties. Therefore, investors would be better off not focusing on detailed estimation and forecasting of parameters but instead on averaging over all possible scenarios or estimating the worst-case scenario. This can be achieved in the EU approach to portfolio allocation. 

In our prior work \cite{Markov:2023}, we examined the portfolio allocation problem with uncertain expected returns for both Gaussian and Asymmetric Laplace-distributed (ALD) returns. Our findings illustrate that uncertainty in expected returns leads to shrinkage in optimal portfolio weights, while skewness and fat tails alter the risk-term dependence. This paper focuses on the portfolio allocation problem involving an uncertain covariance matrix. Our methodology is loosely inspired by the construction of Bayesian models. In this context, the model parameters — expected returns and covariance matrix — are treated as random variables with a specified distribution and corresponding hyperparameters. To obtain observables, we marginalize (integrate) over these random parameters.

The distribution of parameters, which is external to the model, reflects the modeler's perspective on future parameter values and their associated uncertainty. This uncertainty primarily arises from two factors: statistical error due to finite sample estimation and, more significantly, prediction error in a non-stationary environment. Such non-stationarity makes it impossible to precisely estimate future parameter values, even with the most advanced forecasting algorithms. In this paper, we propose three covariance matrix noise models that capture both types of errors in different scenarios.

This paper is structured as follows: First, we will briefly compare the expected utility and mean-variance approaches to portfolio allocation. Second, we will discuss the marginalization over the variance distribution in the univariate case. Third, we compute the expected value of the utility function and corresponding allocation weights by marginalizing the covariance matrix using three analytically solvable noise models: the Wishart distribution, an equivariance model with a block structure, and a two-state model that considers the non-zero probability of a market crash. Finally, we will discuss the practical application of the results.

\section{Mean-variance vs. Expected utility approaches}

In the standard economic approach, asset allocations are derived by maximizing the expected value of the investor's utility function. The utility non-linearly transforms the investor's  wealth and encodes his aversion to risk. The most commonly used utility model in finance is the CARA~\cite{Arrow:1966, Pratt:1964}. The utility functions $U_a(x,a)$ for CARA with respect to the investment outcome (return) $x$ and the risk aversion parameter $a$ can be expressed as follows:
\be
 U_a(x,a) = 
    \begin{cases}
       \frac{1-\exp(-a x)}{a}, &  \mbox{if }a \neq 0 \\
        x, &  \mbox{if }  a = 0
    \end{cases}, \,\, \,
\label{eq:utility}
\ee

To determine the optimal portfolio weights, we maximize the expected value of the utility function $U(x,a)$ with respect to $w$:
\be
w=arg \max_w E_P[U_a(x,a)]=arg \max_w \int dx\,\, U_a(x,a) P(x)
\label{eq:potfolio-opt}
\ee
where $P(x)$ is the distribution of outcome $x$.  

An investor using the MV approach optimizes the following function:
\be
w^* = \arg \max_w [E[x] + \lambda \text{Var}(x)]
\label{eq:mv}
\ee

While the two approaches yield identical results in the case of a normally distributed outcome, they diverge when the outcome distribution cannot be fully characterized by only its first two moments. Obviously, the expected utility approach makes it straightforward to incorporate distributions with skewness and excess kurtosis. Some distributions, like the Asymmetric Laplace Distribution (ALD), offer complete analytical tractability, which is no more complex than that of the Gaussian distribution \cite{Markov:2023}.

If the future covariance matrix $\boldsymbol \Sigma$ and the expected returns $\boldsymbol \mu$ are random variables, the optimal allocation weights are determined by integrating over all possible realizations of the covariance matrix $\boldsymbol \Sigma$ with probability distribution $\boldsymbol P(\Sigma)$ and the expected returns $\boldsymbol \mu$ with probability distribution $P(\boldsymbol \mu)$. The optimal weights $w^*$ are given by:
\be
w^*=\arg \max_w \int d \boldsymbol \Sigma \,\, \int d \boldsymbol \mu \,\, \int d \boldsymbol x\,\,  U_a(x,a) P(x;\boldsymbol \mu,\boldsymbol \Sigma) P(\boldsymbol \mu) P(\boldsymbol \Sigma) 
\label{formula-port}
\ee
where $P(\boldsymbol x)$ is the distribution of portfolio returns $\boldsymbol x$.
$P(\boldsymbol x)$ can be assumed to follow a Gaussian distribution, as the fat tails of the actual return distribution can be attributed to the integration over variance. The integration over $\boldsymbol \Sigma$ is a basic operation in multivariate statistics and involves integrating over all $\frac{N (N+1)}{2}$ independent components of $\boldsymbol \Sigma$. If $P(\boldsymbol \Sigma)$ has a discrete distribution, the integration is substituted by a summation over the corresponding probabilities.

For multivariate Gaussian returns  $r_t \sim N(\boldsymbol \mu,\boldsymbol  \Sigma)$,
the expected utility is given by:
\be
E_N[U_a(\boldsymbol \mu,\boldsymbol \Sigma)] = \int dx\,\, U_a(x,a) P(x;\bs \mu,\bs \Sigma) \sim  (-1) e^{\frac{1}{2}a^2 \boldsymbol w^T \boldsymbol \Sigma \boldsymbol w-a \boldsymbol  \mu^T \boldsymbol  w}
\label{eq:eu1}
\ee

As one can see, the terms related to $\bs \Sigma$ and $\bs \mu$ are separated and can be integrated independently. Correspondingly, we are interested in: 
\be 
\int d \bs \Sigma \,e^{\frac{1}{2}a^2 \boldsymbol w^T \boldsymbol \Sigma \boldsymbol w}  P(\bs \Sigma)=E_{P(\bs \Sigma)}[e^{Tr (\bs W \bs \Sigma)}] = M(\bs W),
\label{eq:eu2}
\ee
where $\bs W=\frac{a^2}{2} \boldsymbol w^T \boldsymbol w$ is the Hadamard product of weights $\bs w$ and  $M(\bs W)$ is the momentum generating function (mgf) for the distribution $P(\bs \Sigma)$. Here, we used the identity $\boldsymbol w^T \boldsymbol \Sigma \boldsymbol w=Tr (\boldsymbol W \boldsymbol \Sigma)$.

The integration over the expected returns $\boldsymbol \mu$ can be done using the following formal analogy:
\be
\int d \boldsymbol \mu \,e^{-a \boldsymbol \mu^T \boldsymbol w}  P(\boldsymbol \mu)=
   E_P[e^{i \boldsymbol t \boldsymbol x}] {\vert_{\boldsymbol t=i a \boldsymbol w }}
\ee 

The last term is a characteristic function (cf) $E_P[e^{i t x}]$ of the probability distribution $P(\mu)$, well known for many statistical distributions.

The goal of this paper is to examine the impact of uncertainties in future variances and correlation coefficients on optimal portfolio allocation. For a general distribution $P(\bs \Sigma)$, the marginalization over $\bs \Sigma$ cannot be done analytically. Therefore, we investigate three analytically tractable models of covariance matrix distribution that may be relevant for practical applications. First, we model $\bs \Sigma$ as being distributed according to the Wishart distribution. Second, we study equivariance block diagonal covariance matrices where the variance of each block is distributed based on the shifted gamma distribution, and the correlation matrix remains fixed. The shift accounts for minimal future volatility, which is an important feature of the stock market. In the third model, we calculate the expected value of the utility function in two scenarios: with probability $p$,  we assume the future market to be in a normal regime with parameters $(\bs \mu_n,\bs \Sigma_n)$, and with probability  $(1-p)$, the market is under stress with parameters $(\bs \mu_s,\bs \Sigma_s)$. This approach allows for the construction of a robust portfolio if a modeler anticipates a non-zero probability of a market crash.

It is important to note that marginalization over covariance matrix  $\bs \Sigma$ produces a trivial result in the MV approach, which is linear in the risk term $R\sim \bs \Sigma$. In this scenario, integration over all possible realizations of  $\bs \Sigma$  yields the expected value of $E[\bs \Sigma]$, and the parameter controlling the uncertainty of $\bs \Sigma$ is absent in the final result. 
Regarding the uncertainty of expected returns, the Black-Litterman model serves as the primary means to account for it in the MV approach \cite{BL:1990}. Despite 30 years of development, the practical use of the model remains limited due to a slew of non-observable and challenging-to-guess parameters. We contend that it is more natural and computationally efficient to directly incorporate future market perspectives by treating both $\bs \mu$ and $\bs \Sigma$ as distributions of random variables and marginalizing (averaging) over their realizations in the EU approach.

\section{Univariate case: Marginalizing over variance and expected return}

In this section, we derive the optimal allocation weight after marginalizing over a one-dimensional covariance matrix, which is the variance. Although this is a simplified case, it offers full analytical tractability of the problem. Many features of this solution can also be extended to the multidimensional case.

\subsection{Shifted gamma distribution as a noise model for variance}

Assume that returns are distributed according to the normal distribution $r_t\sim N(0,\sigma^2)$. Then, the sample variance  $s^2$ of $n$ observations follows the chi-squared distribution $\chi^2_n(\sigma^2)$:
\be
s^2\sim \frac{1}{n} \sum_{i=1}^n r_i^2=\frac{1}{n} \chi^2_n(\sigma^2)
\ee 

The chi-squared distribution $\chi^2_n(\sigma^2)$ is a one-dimensional version of the Wishart distribution, which will be discussed in the subsequent section. Additionally, it has a relationship with the gamma distribution, given by the identity $\chi^2_n(\sigma^2)=\Gamma(\frac{n}{2},2 \sigma^2)$. An important feature of the equity market is the presence of minimal variance  that is not accounted for by the $\chi^2_n(\sigma^2)$ distribution. In other words, variance can be divided into a minimal variance deterministic part and a stochastic part. As a result, we employ a gamma distribution shifted by minimal variance as the model for the future variance distribution.  The shifted gamma distribution is flexible enough to capture both finite sample uncertainty and forecasting power limitations by treating $n$ as a model parameter. 

The probability density function of the three-parameter gamma distribution is given by:
\be
\Gamma(x,\tilde \alpha,\tilde  \beta,\tilde  \gamma)=\frac{(x-\tilde  \gamma)^{\tilde  \alpha-1} e^{-\frac{(x-\tilde  \gamma)}{\tilde  \beta}}}{\tilde  \beta^{\tilde \alpha} \Gamma(\tilde  \alpha)}
, \,\,\,\, x>\tilde  \gamma,\tilde  \alpha>0,\tilde  \beta>0 
\ee

The standard two-parameter (shape-scale) parametrization corresponds to $\tilde \gamma=0$ , and the exponential distribution corresponds to parameters  $\tilde  \alpha=1,\tilde  \gamma=0$. 
The momentum-generating function is given by: 
\be
M_x(t)=E[e^{tx}]=\frac{e^{\tilde  \gamma t}}{(1-\tilde \beta t)^{\tilde \alpha}} 
\label{eq:mgf-gamma}
\ee 

We model the future variance using the following distribution: 
\be
s^2\sim \Gamma(\frac{\alpha}{2}, \frac{2 \sigma^2}{\alpha},\sigma^2_{min}), 
\label{eq:noise-gamma}
\ee
with the shape parameter $\frac{\alpha}{2}$, the scale parameter $\frac{ 2 \sigma^2}{\alpha}$,  and $\sigma^2_{min}$ shifting the distribution by a minimum variance value $\sigma^2_{min}$. 
The mean is given by $E[s^2]=\sigma^2_{min}+\sigma^2$. We use this equation to estimate $\sigma$. The uncertainty parameter $\alpha $ can be estimated from $Var[s^2]=\frac{2 \sigma^4}{\alpha}$. The larger values of $\alpha$ correspond to a lower variance around the target value $E[s^2]$. 

Approximate values of the parameters $\sigma_{\text{min}}$ and $\alpha$ can be estimated from historical market data. The forward-looking 30-day volatility, as measured by the VIX index, reached a minimum value of 9.5 between the years 2006 and 2023. This value can be used to determine the approximate value $ \sigma_{\text{min}} \approx 10\%$. The volatility of volatility is measured by the VVIX Index, which represents the expected volatility (standard deviation) of the 30-day forward price of the VIX Index. The typical value of the VVIX index is in the 80 percent range. This corresponds to a very high level of uncertainty in volatility, with $\alpha < 1$, which makes the mean value of volatility almost useless when applying the two-sigma rule.

\begin{figure}[htbp]
\centering
\includegraphics[width=0.65\linewidth]{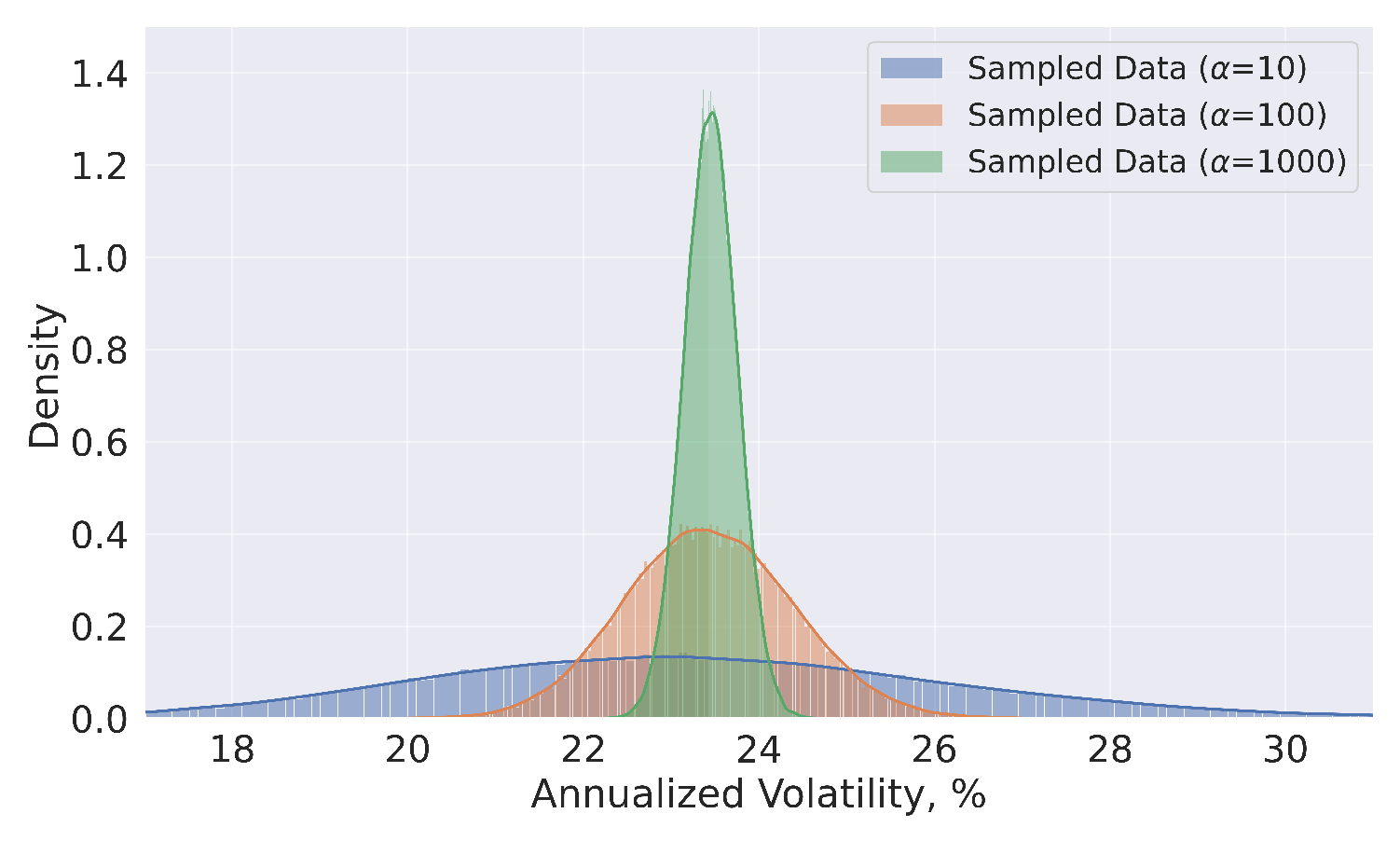}
\caption{Noise model for volatility distribution given by Eq.~\ref{eq:noise-gamma} for different values of $\alpha$}
\label{fig:label_for_all_images}
\end{figure}

In Figure~\ref{fig:label_for_all_images}, we show the distribution of annualized volatility in the model for high, medium, and low uncertainties of  $\sigma^2$ for $\alpha=10$, $\alpha=100$, and $\alpha=1000$, respectively. For this simulation, we set the parameters $\sigma_{min}=0.1$ $(10\%$ annualized) and $\sigma=0.15$ ($15\%$ annualized).  
Subsequently, $N=10^5$ random variances were generated according  to Eq.~\ref{eq:noise-gamma}.  

\subsection{Marginalization over variance and expected return}

To develop intuition, we first discuss the MV framework, which assumes Gaussian returns. Given that the portfolio return $x$ follows a normal distribution $x \sim N(\mu,\sigma)$, the expected utility function $E_N[U_a(x,a)]$ can be expressed as:
\be
E_N[U_a(x,a)]=\int d x\,\, U_a(x,a) N(\mu,\sigma)=\frac{1-e^{\frac{1}{2} a^2 \sigma ^2- a \mu }}{a}
\label{eq:expectedutility}
\ee
where we examine the CARA utility function $U_a(x,a)$ from Eq.~\ref{eq:utility}. According to Eq.~\ref{eq:potfolio-opt}, the optimal portfolio maximizes the expected utility function in Eq.~\ref{eq:expectedutility}. This is equivalent to maximizing the expression $\mu-\frac{a}{2}  \sigma ^2$ or MV in Eq.~\ref{eq:mv}.

The location parameter $\mu$ is often unpredictable. To model it, we assume that $\mu$ follows a normal distribution $\mu \sim N(\mu_0,\sigma_0)$ with parameters $\mu_0$ and $\sigma_0$. We then marginalize (integrate) Eq.~\ref{eq:potfolio-opt} over the location parameter $\mu$. The optimal weight $w^*$ is given by:
\be
w^*=\arg \max_w \int d \mu \,E_N[U_a(x,a)] N(\mu;\mu_0,\sigma_0)=\arg \max_w E_{\mu}[U_a(x,a)]
\label{formula-port-0}
\ee

To derive the optimal weight after marginalization over $\mu$, we rely on the following identity:
\be 
\int_{-\infty}^\infty d\mu\, e^{-a w \mu}  \, N(\mu;\mu_0,\sigma_0)=e^{\frac{1}{2} a^2 \sigma_0^2 w^2-a \mu_0 w}
\label{norm-w-id}
\ee

The optimal weight $w^*_N$ after marginalization over $\mu$ is given by \cite{Markov:2023}:
\be
w^*_{N} = \arg \max_w\left[ (-1) e^{-a (\mu_0-r_0) w+\frac{a^2}{2} w^2 (\sigma^2+\sigma_0^2) }\right]
\label{norm-w-margin}
\ee 

Taking the logarithm of Eq. \ref{norm-w-margin}, we arrive at:  
\be 
w^*_N = \arg \max_w \left[w(\mu_0 - r_0)-\frac{a}{2} w^2 (\sigma^2+\sigma_0^2)\right] = \frac{\mu_0-r_0}{a (\sigma^2+\sigma_0^2)}
\label{mark-w-margin}
\ee
here, $r_0$ is  a risk-free return. In order to avoid trivial cluttering, we assume later in the text that an investor chooses between stock and cash ($r_0$=0). 

We model variance with the model in Eq.~\ref{eq:noise-gamma}:  $s^2\sim \Gamma(\frac{\alpha}{2}, \frac{2\sigma^2}{\alpha},\sigma^2_{min})$. The optimal weight $w^*$ is given by: 
\be
w^*=\arg \max_w \int d s^2 \,\,  E_N[U_a(x,a)] \Gamma(s^2; \frac{\alpha}{2}, \frac{2 \sigma^2}{\alpha},\sigma^2_{min})=\arg \max_w E_{\sigma^2}[U_a(x,a)]
\label{formula-port-1}
\ee

Using the mgf of the gamma distribution in Eq.~\ref{eq:mgf-gamma} and the identity in Eq.~\ref{norm-w-id}, the expected utility function $E_{\mu,\sigma^2}[U_a(x)]$ after marginalization over variance $s^2$ and expected return $\mu$ is given by: 
\be
E_{\mu,\sigma^2}[U_a(x,a)]\sim (-1) e^{\frac{a^2}{2} (\sigma^2_{min}+\sigma_0) w^2-a w \mu_0  }\left(1-\frac{a^2}{\alpha} w^2 \sigma^2 \right)^{-\frac{\alpha}{2}}
\ee

Maximization leads to a cubic equation: 
\be
a^3  (\sigma^2_{min}+\sigma^2_{0}) \sigma^2 w^3- a^2 \mu_0 \sigma^2 w^2- a \alpha (\sigma^2+\sigma^2_{min}+\sigma^2_{0}) w+\mu_0 \alpha=0  
\ee

The solution is a cumbersome expression given by the Cardano formula. The asymptotics for $\mu_0\to 0$ ($\mu_0 \ll \sigma^2_{min}+\sigma^2_{0}+\sigma^2$) is given by:
\be
w^*=\frac{\mu_0}{a(\sigma^2_{min}+\sigma^2_{0}+\sigma^2)}-\frac{\sigma^4 \mu_0^3}{a \alpha (\sigma^2_{min}+\sigma^2_{0}+\sigma^2)^4}+O(\mu_0^4) 
\label{eq:univ-asympt-0}
\ee 

The asymptotics for $\mu_0\to \infty$ ($\mu_0 \gg \sigma^2_{min}+\sigma^2_{0}+\sigma^2$) is given by:
\be
w^*=\frac{\sqrt{\alpha}}{a\sigma}-\frac{\alpha}{2 a \mu_0}+O(\mu_0^{-2}) 
\label{eq:univ-asympt-00}
\ee 

The asymptotics for $\alpha \to \infty$  (small variance uncertainty) is given by:
\be
w^*=\frac{\mu_0}{a(\sigma^2_{min}+\sigma^2_{0}+\sigma^2)}-\frac{\sigma^4 \mu_0^3}{a \alpha (\sigma^2_{min}+\sigma^2_{0}+\sigma^2)^4}+O(\alpha^{-\frac{3}{2}})  
\label{eq:univ-asympt-1}
\ee 

The leading term corresponds to the MV solution but with additional regularization (shrinkage) due to the expected return uncertainty $\sigma_0^2$. We note that an increase in  $\alpha$ (a decrease in variance uncertainty) leads to an increase in the allocation to the risky asset  $w^*$. 

The asymptotics for $\alpha \to 0$  (large variance uncertainty) is given by:
\be
w^*=\frac{\sqrt{\alpha}}{a\sigma}-\frac{\alpha}{2 a \mu_0}+O(\alpha^\frac{3}{2}) 
\label{eq:univ-asympt-2}
\ee 
The leading term corresponds to the inverse volatility allocation. 

The asymptotics for $\sigma^2_{0} \to \infty$  (large expected return uncertainty $\mu_0 \ll \sigma^2_{0}$ ) is given by:
\be
w^*=\frac{\mu_0}{a\sigma^2_{0}}-\frac{\mu_0 (\sigma^2_{min}+\sigma^2)}{a \sigma^4_{0}}+O(\sigma_{0}^{-6})  
\label{eq:univ-asympt-3}
\ee 

The asymptotics for $\sigma^2_{0} \to 0$  and $\sigma_{min}=0$ (small expected return uncertainty $\mu_0 \gg \sigma^2_{0}$) is given by:
\be
w^*=\frac{-\sigma \alpha+\sqrt{\alpha (4 \mu_0^2+\sigma^2 \alpha)}}{2 a \mu_0 \sigma}+O(\sigma^2_{0}) 
\label{eq:univ-asympt-4}
\ee 

We discuss the behavior of this case as a function of  $\alpha$ and the square of the Sharpe ratio  $\frac{\mu_0^2}{\sigma^2}$ in the next section.

\section{Multivariate case: Marginalization over covariance matrices}

The integration over the covariance matrix distribution $P(\bs \Sigma)$ in Eq.~\ref{formula-port} cannot be performed for an arbitrary distribution. Therefore, in this section, we study three models of covariance matrix distributions that allow analytical tractability of the problem.

\subsection{Model 1: Marginalization over Wishart distribution of covariance matrices}

The Wishart distribution $\bs S \sim W_N(n,\bs \Sigma)$ is commonly used to model sample covariance matrices of  multivariate Gaussian data $\bs X\sim N(0,\bs \Sigma)$, where $\boldsymbol  \Sigma \in R^{N \times N}$ \cite{Anderson:2003}. The multivariate distribution can be seen as an extension of the chi-squared distribution, and it naturally emerges when analyzing the sample covariance matrix of multivariate normal data. When we sample $n$ observations from the distribution, the sample covariance matrix $\bs X$ follows a Wishart distribution with a probability density function given by:
\be
f(\bs X|\bs \Sigma,n) = \frac{|\bs X|^{\frac{n}{2}-\frac{N+1}{2}} e^{-\frac{1}{2} \mathrm{tr}(\bs \Sigma^{-1} \bs X)}}{2^{\frac{n}{2} N} |\bs \Sigma|^{\frac{n}{2}} \Gamma_N\left(\frac{n}{2}\right)} 
\ee
where $|\bs X|$ is the determinant of matrix $\bs X$ and  $\Gamma_N\left(\alpha\right)$ is multivariate gamma function. 

The expected value is $E[\bs X]=n \bs \Sigma$, and the momentum-generating function is $M_W(\bs \Sigma)=E[e^{tr(\bs W \bs \Sigma)}]=|\bs I-2 \bs W \bs \Sigma|^{-\frac{n}{2}}$. Here, $\bs \Sigma$ is an empirical covariance matrix derived from historical data, and $\bs I$ is an identity matrix.

Consequently, we study the following noise model for the covariance matrix $\Sigma$:  
\be 
\bs S \sim W_N(\alpha,\frac{\bs \Sigma}{\alpha})
\label{eq:noise_m1}
\ee
where $\alpha$ is a parameter that controls the noise level. For large $\alpha$ values, most of the sample matrices  $\bs S$ will closely resemble $\bs \Sigma$, while for small $\alpha$ values, the elements of the sample matrices $\bs S$ can deviate significantly from the desired covariance matrix $\bs \Sigma$. The expected value of $\bs S$ is given by the target covariance matrix $E[\bs S]=\bs \Sigma$, and the momentum-generating function is given by:
\be 
M_W(\bs \Sigma)=|\bs I-\frac{2}{\alpha} \bs W \bs \Sigma|^{-\frac{\alpha}{2}}
\ee

In Appendix B, we present the distribution of variances and correlation coefficients of two-dimensional matrices, as described by the model in Eq.~\ref{eq:noise_m1}, for various values of the parameter $\alpha$.

\subsection{The expected utility function and optimal weights}

In the case of multivariate Gaussian returns $\boldsymbol r_t \sim N(\boldsymbol \mu,\boldsymbol  \Sigma)$, the expected utility function to be maximized is proportional to:
\be
E_N[U_a(\bs \mu,\bs \Sigma)]\sim (-1) e^{\frac{a^2}{2} \boldsymbol w^T \boldsymbol \Sigma \boldsymbol w -a \boldsymbol  \mu^T \boldsymbol  w}
\label{opt-markovitz1}
\ee

Correspondingly, the optimization problem to maximize $E_N[U_a(\bs \mu,\bs \Sigma)]$ is given by:
\be
 \boldsymbol  w^*=\arg \max_{ \boldsymbol w} \left[ \boldsymbol  \mu^T \boldsymbol  w-\frac{a}{2} \boldsymbol w^T \boldsymbol \Sigma \boldsymbol w \right],
\label{opt-markovitz2}
\ee
which has the following classical MV solution: 
\be 
 \boldsymbol  w^*=\frac{1}{a} \boldsymbol  \Sigma^{-1} \boldsymbol  \mu 
\ee
The effect of the transaction cost can also be taken into account\footnote{Additionally, one can take into account the transaction costs by including a term proportional to the turnover, $e^{\frac{\eta}{2} (\boldsymbol w-\boldsymbol w_0)^2}$, in the utility function in Eq.~\ref{opt-markovitz1}. Here, $\boldsymbol w_0$ represents the target weights. Since the form is quadratic in $\boldsymbol w$, it leads to a redefinition of $\boldsymbol \mu$ and $\boldsymbol \Sigma$:
$$
\boldsymbol \mu'=\boldsymbol \mu+ \frac{\eta}{a} \boldsymbol w_0 , \,\, \boldsymbol \Sigma'=\boldsymbol \Sigma+ \frac{\eta}{a^2} \boldsymbol I
$$
}. The minimum variance portfolios with normalized weights can be obtained from MV portfolios Eq.~\ref{opt-markovitz2} by setting all elements of $\boldsymbol{\mu}$ to 1. In this case, the term $\boldsymbol{\mu}^T \boldsymbol{w}$ becomes a constant, since the sum of the weights $ \sum_{i=1}^n w_i$ is constant if the weights are normalized. This constant term is irrelevant for optimization.

Marginalization of the expected utility function $E_N[U_a(\bs \mu,\bs\Sigma)]$ over the expected returns $\boldsymbol \mu$, which follow a normal distribution  $\boldsymbol\mu\sim N (\boldsymbol \mu_0,\boldsymbol \Sigma_0)$, can be done analytically:
\be
\int_{-\infty}^\infty d \boldsymbol \mu\, e^{-a \boldsymbol  w^T \boldsymbol  \mu}  \, N(\boldsymbol \mu; \boldsymbol \mu_0,\boldsymbol \Sigma_0)=e^{\frac{1}{2} a^2 \boldsymbol w^T \boldsymbol \Sigma_0 \boldsymbol w-a \boldsymbol \mu_0^T \boldsymbol w}
\ee
where $\boldsymbol \Sigma_0$ is a diagonal matrix with elements equal to the variances of the individual components of $\boldsymbol \mu$. 
  
Integration over all possible realizations of $ \boldsymbol \Sigma$, as given by the noise model in Eq. \ref{eq:noise_m1}, of the expected utility $E_N[U_a(\boldsymbol  \mu,\boldsymbol  \Sigma)]$  in Eq. \ref{opt-markovitz1}, is provided by:
\be
E_{\boldsymbol \Sigma}[U_a(\boldsymbol \mu,\boldsymbol \Sigma)] =\int d \boldsymbol S\,  E_N[U_a(\boldsymbol \mu,\boldsymbol S)] f(\boldsymbol S|\alpha, \frac{\boldsymbol  \Sigma}{\alpha})  
\sim (-1) e^{-a \boldsymbol  \mu^T \boldsymbol  w} |\boldsymbol  I-\frac{a^2}{\alpha} \boldsymbol  W \boldsymbol  \Sigma |^{-\frac{\alpha}{2}} 
\ee

Using the identity $\ln \det (\boldsymbol  X)=Tr \ln (\boldsymbol  X)$, the definition of matrix logarithm $\ln (\boldsymbol I-\boldsymbol X)= \sum_{n=1}^\infty (\boldsymbol I-\boldsymbol X)^n/n$,  and $ (\boldsymbol w^T \boldsymbol \Sigma \boldsymbol w) =Tr ((\boldsymbol w^T \boldsymbol w) \boldsymbol \Sigma)$,  we obtain the expected utility after marginalization over random covariance matrices $S$: 
\be 
E_{\Sigma}[U(\boldsymbol \mu,\boldsymbol \Sigma)]  
\sim (-1)e^{-a \boldsymbol  \mu^T \boldsymbol  w-\frac{\alpha}{2} \ln \left[1-\frac{a^2}{\alpha}  (\boldsymbol w^T \boldsymbol \Sigma \boldsymbol w)\right]}
\label{normal-eu-1} 
\ee

Combining this result with marginalization over the expected returns  $ \boldsymbol \mu$, we arrive at:
\be 
E_{\boldsymbol \mu,\boldsymbol \Sigma}[U(\boldsymbol \mu,\boldsymbol \Sigma)]  
\sim (-1)e^{-a \boldsymbol  \mu_0^T \boldsymbol  w+\frac{a^2}{2}  \boldsymbol w^T \boldsymbol  \Sigma_0 \boldsymbol w-\frac{\alpha}{2} \ln \left[1-\frac{a^2}{\alpha}  (\boldsymbol w^T \boldsymbol \Sigma \boldsymbol w)\right]}
\label{normal-eu-1a} 
\ee

Later in the text, we assume that the parameters $a$ and $\alpha$ are set in such a way that the logarithm is real in order to avoid unnecessary notation clutter.

The optimal weights $\boldsymbol  w^*$, considering the uncertain expected return and covariance matrix, are given by:
\be
\boldsymbol w^*=\arg \max_{\boldsymbol w} \left[  \boldsymbol \mu_0^T \boldsymbol w-\frac{a}{2}  \boldsymbol w^T \boldsymbol  \Sigma_0 \boldsymbol w +\frac{\alpha}{2 a} \ln \left[1-\frac{a^2}{\alpha}  (\boldsymbol w^T \boldsymbol  \Sigma \boldsymbol w)\right] \right]
\label{eq:opt-weight1}
\ee
The optimization problem is convex and can be solved with constraints by a numerical optimizer. 
In the limit of a large $\alpha$  (representing a small noise level in the covariance matrix  $\bs \Sigma$), the Taylor expansion of Eq.~\ref{eq:opt-weight1} can be expressed as:
\be
\boldsymbol w^*=\arg \max_{\boldsymbol w} \left[  \boldsymbol \mu_0^T \boldsymbol w-\frac{a}{2}  \boldsymbol w^T  (\boldsymbol  \Sigma+\boldsymbol  \Sigma_0) \boldsymbol w \right]
\label{eq:opt-weight2}
\ee

The solution is given by:
\be
\boldsymbol w^*=\frac{1}{a}(\boldsymbol \Sigma+\boldsymbol \Sigma_0)^{-1}\boldsymbol \mu_0 
\label{eqsolution1norm}
\ee

For the case of zero uncertainty in the expected returns, with $\bs \Sigma_0=0$, we have the following optimization problem:
\be
\boldsymbol w^*=\arg \max_{\boldsymbol w} \left[  \boldsymbol \mu_0^T \boldsymbol w+\frac{\alpha}{2 a} \ln \left[1-\frac{a^2}{ \alpha}  (\boldsymbol w^T \Sigma \boldsymbol w)\right] \right]
\label{eq:opt-weight3}
\ee

Maximizing Eq.~\ref{eq:opt-weight3} yields the following system of quadratic equations with respect to $\boldsymbol w$: 
\be
\boldsymbol \mu_0  \left (1 -\frac{a^2}{\alpha} \boldsymbol  w^T \boldsymbol \Sigma \boldsymbol w \right)-a \boldsymbol \Sigma \boldsymbol w=0 
\label{eq:ald-q2}
\ee

The solution is given by:
\begin{equation}
\boldsymbol w^* = \frac{1 - \frac{a^2}{\alpha} d}{a} \boldsymbol \Sigma^{-1} \boldsymbol \mu_0=g_W(q,\alpha) \times \left(\frac{1}{a}\boldsymbol \Sigma^{-1} \boldsymbol \mu_0 \right)
\label{eq:ald-q3}
\end{equation}
where $d = \boldsymbol w^T \boldsymbol \Sigma \boldsymbol w$, $q = \boldsymbol \mu_0^T \boldsymbol \Sigma^{-1} \boldsymbol \mu_0$, and $g_W(q,\alpha)=1 - \frac{a^2}{\alpha} d$. By substituting the solution \eqref{eq:ald-q3} into the definition of $d$, we obtain a quadratic equation with the following solution:
\begin{equation}
d = \frac{\alpha}{2 a^2 q} \left(2q+\alpha-\sqrt{\alpha(4 q+\alpha)}\right)
\end{equation}
The solution provided by Eq.~\ref{eq:ald-q3} coincides with Eq.~\ref{eq:univ-asympt-4} in the univariate case.

We note that the solution in Eq.~\ref{eq:ald-q3} appears similar to the optimal weights calculated in \cite{Markov:2023} for outcomes that follow a Laplace distribution:
\begin{equation}
\boldsymbol w_{LD}^* = g_{LD}(q)\times \left(\frac{1}{a} \boldsymbol \Sigma^{-1} \boldsymbol \mu_0\right)
\label{eq:ald-q4}
\end{equation}
where 
\be 
g_{LD}(q)=1 - \frac{a^2 \tilde d}{2}, \,\,  \tilde d = \frac{2 (1 + q - \sqrt{1 + 2 q})}{a^2 q}
\ee

\begin{figure}[t]
    \centering
    \includegraphics[width=\linewidth]{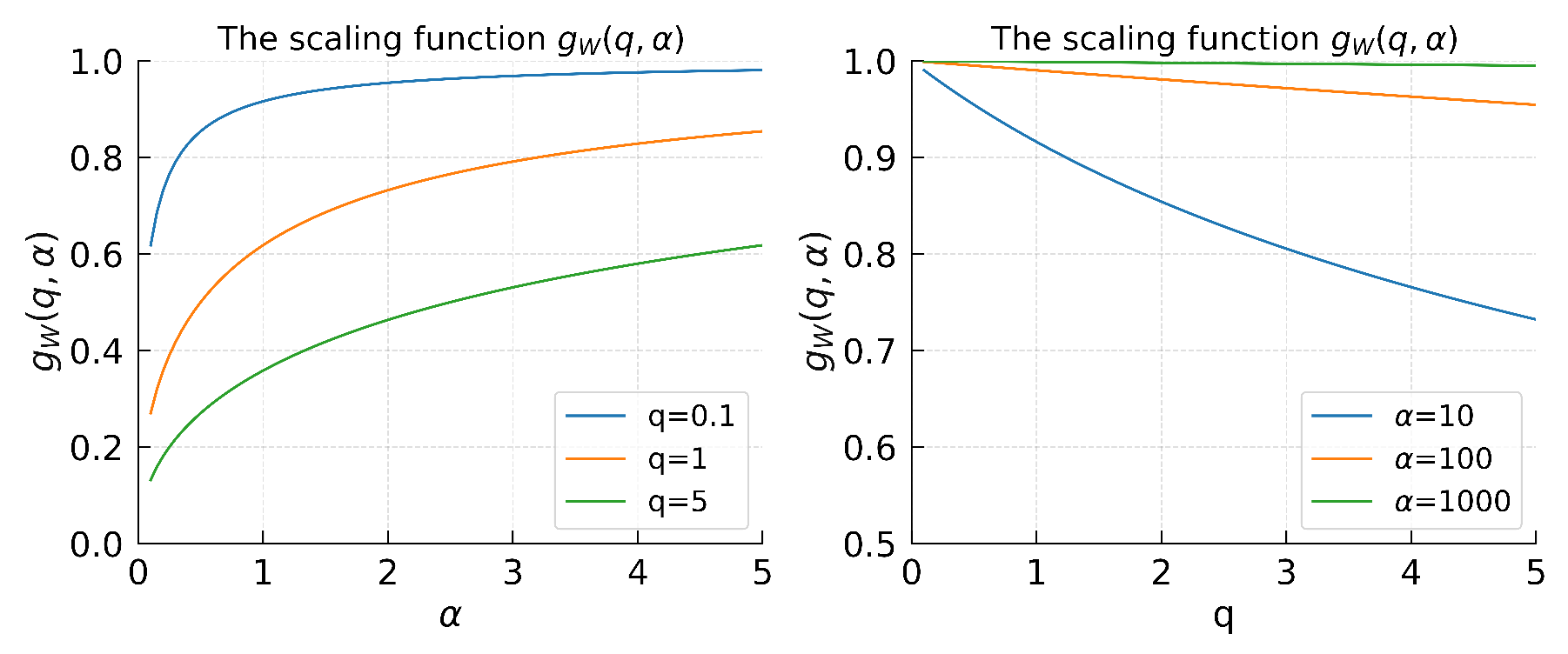}
    \caption{Scaling function $g_W(q,\alpha)$}
    \label{fig:my_label}
\end{figure}

The scaling function $0<g_W(q,\alpha)\le 1$ does not depend on the risk aversion parameter $a$. It represents the decrease in the absolute allocation level due to non-zero uncertainty regarding the future covariance matrix in comparison to the standard MV solution $\boldsymbol w^* =\frac{1}{a} \boldsymbol \Sigma^{-1} \boldsymbol \mu$. Thus, $g_W(q,\alpha)$  can be used to compare allocation levels for different return-to-risk parameters $q$ and varying levels of covariance matrix noise $\alpha$. 
We present the function $g_W(q,\alpha)$ in Figure~\ref{fig:my_label} for different values of $q$ and $\alpha$. The scaling function $g_W(q,\alpha)$ is positive and decreases as the noise level $\alpha$ increases. This suggests that as uncertainty about future $\bs \Sigma$ decreases, the allocation level also decreases. In the conventional MV approach, weights are normalized to one $\sum_{i=1}^{N} w_{i} = 1$, and only relative weights are considered. While using the EU approach, we provide insights into both the absolute level and relative weights as functions of model parameters due to the nature of absolute risk in the utility function. In other words, accounting for uncertainty in the covariance results in a decrease in the absolute level while maintaining the same relative allocation.

If the uncertainty of the expected returns is not zero $\bs \Sigma_0\neq 0$, the optimal weights $\boldsymbol w^*$ are given by:
\be
\boldsymbol w^*=\arg \max_{\boldsymbol w} \left[  \boldsymbol \mu_0^T \boldsymbol w-\frac{a}{2}  \boldsymbol w^T \boldsymbol  \Sigma_0 \boldsymbol w +\frac{\alpha}{2 a} \ln \left[1-\frac{a^2}{\alpha}  \boldsymbol w^T \bs \Sigma \boldsymbol w \right]\right]
\label{eq:opt-weight4}
\ee

Eq.~\ref{eq:opt-weight4} can be solved by a convex numerical optimizer, such as CVXPY~\cite{Diamond:2016}, even when additional constraints, such as long-only, maximum number of positions, turnover, and other constraints, are introduced. The unconstrained solution of Eq.~\ref{eq:opt-weight4} can be obtained by taking the derivative with respect to $\boldsymbol w$. Thus, we arrive at the following system of cubic equations:
\be
\boldsymbol (\boldsymbol \mu_0-a \boldsymbol \Sigma_0 \boldsymbol w)  \left (1 -\frac{a^2}{\alpha} \boldsymbol  w^T \boldsymbol \Sigma \boldsymbol w \right)-a \boldsymbol \Sigma \boldsymbol w=0 
\label{eqopt-mald}
\ee

The solution is given by:
\begin{equation}
\boldsymbol w^* = \frac{(1-\frac{a^2}{\alpha} d) \boldsymbol \mu_0}{a(  \boldsymbol \Sigma+(1-\frac{a^2}{\alpha} d) \boldsymbol \Sigma_0)}
\label{eq:ald-marg-q3}
\end{equation}
where $d = \boldsymbol w^T \boldsymbol \Sigma \boldsymbol w$. By substituting the solution from Eq.~\ref{eq:ald-marg-q3} into the definition of $d$, we obtain a nonlinear equation for $d$ that can be solved numerically. It is worth noting that the uncertainty $\boldsymbol \Sigma_0$ of the $\boldsymbol \mu$ plays a similar role to the regularization (shrinkage) approach detailed in \cite{Ledoit:2003}. However, the EU approach provides a clearer interpretation, and there is no need to estimate a shrinkage constant.

In the case of non-zero uncertainty in expected returns, i.e., $\boldsymbol \Sigma_0 \neq 0$, both the analytical solution in Eq.~\ref{eq:ald-marg-q3} and the numerical simulations indicate that the solution of Eq.~\ref{eq:opt-weight4} can deviate (sometimes significantly) from the MV solution $\boldsymbol w_{MV}^*=\frac{1}{a} \boldsymbol \Sigma^{-1} \boldsymbol \mu_0$. In this case, the uncertainties of expected returns and the covariance matrix interfere with each other.

\subsection{Model 2: Marginalization over variance using block covariance matrices}

In \cite{Markov:2023}, we proposed deriving the covariance matrix from a precision matrix with a graphical structure determined by conditional independence. The resulting matrix has a block structure. The graphical structure reduces the number of nonzero elements in the covariance matrix, thereby improving the condition number and making the matrix inversion operation more stable. In this section, we investigate allocation rules for two analytically solvable cases of covariance matrices, $\bs \Sigma_{m1}$ and $\bs \Sigma_{m2}$. In the first model, each block has an equivariance structure, but there are no cross-correlation terms between blocks:
\begin{equation}
\bs \Sigma_{m1}=\begin{bmatrix}
(\sigma_{min;1}^2+\sigma_1^2) R_1 & & & \\
& (\sigma_{min;2}^2+\sigma_2^2) R_2 & & \\
& & \ddots & \\
& & & (\sigma_{min;K}^2+\sigma_K^2) R_K \\
\end{bmatrix},
\end{equation}

In the second model, all blocks have the same variance, but the correlation structure is encoded in equicorrelation blocks: 
\begin{equation}
\bs \Sigma_{m2}=(\sigma_{min}^2+\sigma^2)  R_{2m}=(\sigma_{min}^2+\sigma^2) \begin{bmatrix}
R_{11} & R_{12} & \cdots & R_{1K} \\
R_{21} & R_{22} & \cdots & R_{2K} \\
\vdots & \vdots & \ddots & \vdots \\
R_{K1} & R_{K2} & \cdots & R_{KK} \\
\end{bmatrix}
\end{equation}

We marginalize the variance of each block, assuming that the future variance follows a shifted gamma distribution.

If a covariance matrix $\Sigma$ has a block equivariant structure, the bilinear form $\boldsymbol w \boldsymbol \Sigma_{1m} \boldsymbol w$ is given by the sum over $K$ blocks:
\be 
\boldsymbol w \boldsymbol \Sigma_{m1} \boldsymbol w=\sum_{i=1}^K \sigma^2_i   \boldsymbol w_i \boldsymbol R_i \boldsymbol w_i
\ee
where $\sigma^2_i$, $\boldsymbol R_i$, and  $\boldsymbol w_i$ represent the $i$-th block variance, correlation matrix, and weights, respectively. Assuming that each block variance $\sigma_i$ follows a shifted gamma distribution, we derive the expected utility function by marginalizing over variance and expected returns:
\be
E_{\boldsymbol \mu,\boldsymbol \Sigma}[U(x)]\sim (-1) e^{-a \boldsymbol w^T \boldsymbol \mu+ \frac{a^2}{2} \sum_{i=1}^K \boldsymbol \sigma^2_{min;i} \boldsymbol  w_i^T \boldsymbol R_i \boldsymbol  w_i }\prod_{i=1}^K \left(1-\frac{a^2 \sigma_i^2}{\alpha_i} \boldsymbol w_i \boldsymbol R_i \boldsymbol w_i \right)^{-\frac{\alpha_i}{2}}
\label{eq:opt-b0}
\ee

This leads to the following optimization problem: 
\be
\boldsymbol w^*=\arg \max_{\boldsymbol w} \left[\boldsymbol \mu_0^T \boldsymbol w-\frac{a}{2}  \boldsymbol w^T \boldsymbol  \Sigma_0\boldsymbol w- \frac{a}{2}  \sum_{i=1}^K \boldsymbol \sigma^2_{min;i} \boldsymbol  w_i^T R_i \boldsymbol  w_i +\sum_{i=1}^K \frac{\alpha_i}{2 a} \ln\left[1-\frac{a^2 \sigma_i^2}{\alpha_i}  \boldsymbol w_i^T \boldsymbol R_i \boldsymbol w_i\right] \right]
\label{eq:opt-b1}
\ee

Eq.~\ref{eq:opt-b1} can be solved numerically using a convex optimizer, with additional constraints if needed. In the limit of large $\alpha$ (small noise level), we have the standard MV accompanied by shrinkage:
\be
\boldsymbol w^*=\arg \max_{\boldsymbol w} \left[  \boldsymbol \mu_0^T \boldsymbol w-\frac{a}{2}  \boldsymbol w^T  (\boldsymbol  \Sigma_{min}+\boldsymbol  \Sigma+\boldsymbol  \Sigma_0) \boldsymbol w \right]
\label{eq:opt-b2}
\ee
where $ \boldsymbol w^T \bs  \Sigma  \boldsymbol w= \sum_{i=1}^K  \sigma^2_{i} \boldsymbol w_i^T \bs R_i \boldsymbol w_i$ and  $ \boldsymbol w^T \bs \Sigma_{min} \boldsymbol w= \sum_{i=1}^K \sigma^2_{min;i} \boldsymbol  w_i^T \bs R_i \boldsymbol  w_i$.  

In the second model, each block $\bs R_{ij}$ of the covariance matrix $\bs \Sigma_{2m}$ is represented by an equicorrelation matrix: 
\begin{equation}
R_{ij} = 
\begin{bmatrix}
1 & \rho_{ij} & \cdots & \rho_{ij} \\
\rho_{ij} & 1 & \cdots & \rho_{ij} \\
\vdots & \vdots & \ddots & \vdots \\
\rho_{ij} & \rho_{ij} & \cdots & 1 \\
\end{bmatrix}
\end{equation}

At first glance, the model $\bs \Sigma_{2m}$ might seem like a crude approximation. However, in practice, sample estimates may not align with the population parameters of the covariance matrix, making precise modeling unwarranted. Furthermore, instead of the $N(N+1)/2$ parameters in the original covariance matrix, the matrix $\bs \Sigma_{2m}$ requires fewer parameters to estimate. Specifically, it only requires a single univariate variance $\sigma^2$ and $K(K+1)/2$ correlation coefficients $\rho_{ij}$, which can be taken as averages per block \cite{Corey:1998}. The block assumption ensures that the matrix $\bs \Sigma_{2m}$ is invertible. In \cite{Archakov:2022}, advanced statistical methods were introduced to estimate matrices with such block structures. The global industry classification standard (GICS) groups emerged as the best model for block selection based on minimizing the Bayesian Information Criterion.

Thus, the optimization problem is as follows: 
\be
\boldsymbol w^*=\arg \max_{\boldsymbol w} \left[\boldsymbol \mu_0^T \boldsymbol w-\frac{a}{2}  \boldsymbol w^T \boldsymbol  \Sigma_0\boldsymbol w- \frac{a}{2}  \sigma^2_{min} \boldsymbol  w^T \bs R_{2m} \boldsymbol  w + \frac{\alpha}{2 a} \ln\left[1-\frac{a^2 \sigma^2}{\alpha}  \boldsymbol w^T \boldsymbol R_{2m} \boldsymbol w\right] \right]
\label{eq:opt-b3}
\ee

Based on the symmetry of the problem (assuming that the expected return parameters $\bs \mu_0$ and $\bs \Sigma_0$ follow the same block structure), stocks within each $i$th block have equal weight allocation:
\be
\bs w_i=[w_i,w_i,\cdots,w_i]
\ee
and the optimization in Eq.~\ref{eq:opt-b3} only needs to find $K$ block weights $w_i$, leading to more stable weights.

\subsection{Model 3: Two-state scenario optimization}
Stock markets do not grow in a linear trend. There are multiple periods when (almost) all assets experience severe and prolonged declines. In this section, we aim to calculate portfolio allocation rules for situations where a modeler assigns a non-zero probability of a market crash or correction occurring during the expected holding period of the portfolio.

Financial correlation matrices have a unique property: the correlation between stocks increases with volatility. In the event of a market crash, volatility spikes, and almost all correlations approach one. Although it is possible to integrate the correlation coefficient over an appropriate distribution, operating in a discrete space is more practical. In this approach, the future is modeled as a two-state system: a normal regime with parameters $\boldsymbol \mu_n$ and $\boldsymbol \Sigma_n$, and a stressed regime with parameters $\boldsymbol \mu_s$ and $\boldsymbol \Sigma_s$. Within this context, the expected value of the utility function is a discrete sum, with probability $p$ assigned to the normal regime and probability $1-p$ to the stressed regime. Consequently, the expected utility of the two-state system $E_{2s}[U_a]$ is given by:
\be
E_{2s}[U_a]=p E[U_a(\boldsymbol \mu_n,\boldsymbol  \Sigma_n)]+(1-p) E[U_a(\boldsymbol  \mu_s,\boldsymbol  \Sigma_s)]  
\label{eq:two-state-1}
\ee
For conceptual clarity, we use the expected utility from Eq.\ref{opt-markovitz1} instead of Eq.\ref{normal-eu-1a}. Taking the logarithm of Eq.~\ref{eq:two-state-1}, we have the following optimization problem: 
\be
\bs w^*= \arg \min_{\boldsymbol w} \left[ \log(\exp{u_n}+\exp{u_s}) \right]
\label{eq:two-state-2}
\ee
where $u_n=\log(p)+ \frac{a^2}{2} \bs w^T \bs \Sigma_n \bs w-a \bs \mu_n \bs w$ and  $u_s=\log(1-p)+ \frac{a^2}{2} \bs w^T \bs \Sigma_s \bs w-a \bs \mu_s \bs w$. 
The optimization problem corresponds to the minimization of the LogSumExp (LSE) function \cite{LSE:wiki} and is convex. This convexity can be inferred from the fact that both \(u_n\) and \(u_s\) are convex functions of $\boldsymbol{w}$, and the LogSumExp function is also convex. LogSumExp is an approximation to the maximum \( \max_{i \in \{n,s\}} u_{i} \) with the following bounds:
\be
\max \left\{u_n,u_s\right\}\le  \log(\exp{u_n}+\exp{u_s}) \le  \max \left\{ u_n,u_s\right \}+\log(2)
\ee
In the limit of high risk aversion parameter $a \to \infty$, the equation Eq.~\ref{eq:two-state-2} simplifies to: 
\be
\bs w^*= \arg \min_{\boldsymbol w} \max_{n,s} \left[\frac{a}{2} \bs w^T \bs \Sigma_n \bs w-\bs \mu_n \bs w , \frac{a}{2} \bs w^T \bs \Sigma_s \bs w-\bs \mu_s \bs w \right]
\label{eq:two-state-3}
\ee
Thus, we end up optimizing the MV utility function for the corresponding state. 

In the limit of low risk aversion parameter $a\to 0$, the exponent in the expected value of $\log E_{2s}[U_a]$ can be approximated by the Taylor expansion $e^{x} \approx 1+x+\frac{x^2}{2}$  with respect to the risk aversion parameter $a$ and we have:
\be
\bs w^*= \arg \min_{\boldsymbol w} \left[\frac{a}{2} \bs w^T  \tilde {\bs \Sigma} \bs w-\tilde {\bs \mu^T} \bs w \right] 
\label{eq:two-state-4}
\ee
where $\tilde {\bs \mu}=p \bs  \mu_n +(1-p) \bs \mu_s$ and $\tilde {\bs \Sigma}= p \bs \Sigma_n +(1-p) \bs \Sigma_s+p(1-p)(\bs \mu_n-\bs \mu_s)^T (\bs \mu_n-\bs \mu_s)$. 
Thus $\bs w^*=\frac{1}{a} \tilde {\bs \Sigma}^{-1}\tilde {\bs  \mu}$. In one dimensional case, the optimal weight is given by:
\be
w^*=\frac{1}{a} \frac{p \mu_n +(1-p)\mu_s}{p \sigma_n^2+(1-p) \sigma_s^2+p(1-p) (\mu_n-\mu_s)^2}
\ee 

The minimum variance portfolio with additional constraints $w>0$ and $\sum_{i=1}^N w_i=1$ can be obtained from Eq.~\ref{eq:two-state-4} by setting all expected returns as equal ($\mu_{n;i}=const_1$ and $\mu_{s;i}=const_2$). In this scenario, both $\boldsymbol \mu_n^T \boldsymbol w$ and $\mu_s^T \boldsymbol w$  become constants that are irrelevant for optimization. The optimal weights are then given by: 
\be
\bs w^*=\arg \min_{\boldsymbol w} \left[\bs w^T \tilde {\bs \Sigma} \bs w+c \bs w^T \bs w \right],\,\,\,\, \tilde  {\bs \Sigma} =  p \bs \Sigma_n+(1-p)  \boldsymbol  \Sigma_s   
\label{eq:two-state-6}
\ee

We observe that the resulting covariance matrix \(\tilde{\bs \Sigma}\) is shrunk towards the covariance matrix in the stressed regime \(\bs \Sigma_s\) with a shrinkage coefficient of \(1-p\) with added $L^2$ regularization term $\bs w^T \bs w$. A similar problem involving the CARA utility function and a mixture of multivariate Gaussian returns was also studied in \cite{Boyd:2022}.

In theory, one can extend the formalism to an arbitrary number of market states to better capture the fat tails and skewness of real returns. However, in practice, we need to model future returns and the covariance matrix, both of which are known with significant uncertainty. In such cases, a two-state model appears optimal, and the hypothetical stressed state parameters  $\bs \mu_s,\bs \Sigma_s$
should be modeled using the simplest model possible.

Often, there is no particular benefit in  in precisely modeling the covariance matrix in the stressed regime, $\bs \Sigma_{s}$. For long-only equity strategies,  this matrix  can be represented in a simplified form as an equicorrelation and equivariance matrix with a high correlation coefficient value, $\rho_s \approx 0.8$. Its variance is denoted by $\sigma^2_{s}$. The matrix is defined as:
\be
\boldsymbol \Sigma_{s} = \sigma^2_{s} ((1-\rho_s) \boldsymbol I  + \rho_s \boldsymbol {1}\boldsymbol {1}^T)
\ee
where $\boldsymbol {I}$ is an identity matrix and $\boldsymbol {1}\boldsymbol {1}^T$ is a matrix of ones. In Appendix C, we provide a scatter plot of monthly average volatility vs. monthly average correlation between index constituents for the years 2016 and 2022, which correspond to the normal regime, and for the years 2008 and 2020, which correspond to the stressed regime. The historical values can be used to specify $\bs \Sigma_s$.  

In general, there are three regimes for correlation coefficients in the stressed regime, and the specifics depend on asset classes and investment strategies. In equities, the correlation between assets tends to approach \(\rho_s=1\) as all assets move in unison. For statistical arbitrage or relative value strategies, assets that are typically correlated can become anti-correlated, with \(\rho_s=-1\). Furthermore, during a crash, the correlation between different asset classes might approach zero, \(\rho_s=0\), as price dynamics become chaotic. In such scenarios, price movements are largely influenced by liquidity demands or the need to cover margin calls.

\section{Concluding Remarks}

In both our previous work \cite{Markov:2023} and in this paper, we have investigated possible ways to go beyond the limitations of the classical MV  approach. We found it useful to think of MV as a limiting case of the more general CARA expected utility maximization in the case of multivariate Gaussian returns with zero uncertainty of expected returns and the covariance matrix. We demonstrated that it is often technically easier to start the analysis by utilizing the formal definition of expected utility and its optimization.

In the EU approach, the expected value of the utility function can be calculated for returns distributed according to the ALD~\cite{Markov:2023}. This allows for studying the effects of fat tails and skewness in the outcome distribution. We have shown that stock returns on daily, weekly, and monthly scales can be well approximated by the ALD with moderate skewness values. The skewness of the outcome becomes more important for quantitative strategies, like those with trend-following programs, which demonstrate positive skewness, or options or volatility selling, which exhibit negative skewness. Paradoxically, the expected value of the utility function with returns following a skewed normal distribution cannot be expressed as an elementary function. In this sense, the ALD offers the best of both worlds: it takes into account the properties of real-world return distributions and allows for full analytical tractability of the problem.

Uncertainty about the future values of model parameters is a cornerstone assumption in any financial modeling. To account for this, we model the expected returns and the covariance matrix as random variables. The corresponding expected values of the utility function are obtained by marginalization (integration) over all possible values of these random variables. In simple terms, the marginalization of expected returns results in the shrinkage of the covariance matrix. Taking into account the uncertainty of the covariance matrix leads to a decrease in the absolute level of allocation. This fact is missing when returns are normalized. 

Finally, we demonstrate that the worst-case scenario allows for an analytical solution using the Karush–Kuhn–Tucker (KKT) method \cite{Markov:2023}. In the worst-case scenario, the maximum weight is allocated to the worst-performing asset. The expected utility of such a minimax (MM) portfolio has a convex form: 
\be
\arg \min_w  \left[|\bs w|_{\infty} + \frac{b}{2} \,\, \boldsymbol w^T \Sigma \boldsymbol w\right], \,\, s.t. \sum_{i=1}^N w_i=1, \,w_i\ge 0
\label{mm-min-worst}
\ee
where the infinity norm $|w|_{\infty}$ selects the maximum weight,
defined as $|w|_{\infty} =\max_i \boldsymbol w $. The solution is a combination of uniform and MV weights and does not suffer from the excessive concentration that is typical of the pure MV approach. The utility serves as an alternative to risk-parity, but it is convex and has a clear interpretation.

The diversity of trading styles, asset classes, and the use of leverage by portfolio managers justify the variety of objective functions available for optimization studied in this paper. Also, the exponential utility function is commonly used in many areas of economics and decision-making science. Extensions of MV to skewed, fat-tailed distributions (such as ALD), differentiating between mean optimization and worst-case optimization, and considerations of the uncertainty of outcome distribution parameters are recurring themes in this research. We hope that the results obtained will prove useful  in both theoretical and practical contexts.


\begin{thebibliography}{14}
\providecommand{\natexlab}[1]{#1}
\providecommand{\url}[1]{\texttt{#1}}
\expandafter\ifx\csname urlstyle\endcsname\relax
  \providecommand{\doi}[1]{doi: #1}\else
  \providecommand{\doi}{doi: \begingroup \urlstyle{rm}\Url}\fi

\bibitem[Sharpe(2007)]{Sharpe:2007}
William~F. Sharpe.
\newblock Expected utility asset allocation.
\newblock \emph{Financial Analysts Journal}, 63\penalty0 (5):\penalty0 18--30,
  2007.

\bibitem[Markowitz(1952)]{Markowitz:1952}
Harry Markowitz.
\newblock Portfolio selection.
\newblock \emph{The Journal of Finance}, 7\penalty0 (1):\penalty0 77--91, 1952.

\bibitem[Markov and Markov(2023)]{Markov:2023}
Maxime Markov and Vladimir Markov.
\newblock Portfolio optimization rules beyond the mean-variance approach.
\newblock \emph{arXiv:2305.10403}, 2023.
\newblock URL \url{https://doi.org/10.48550/arXiv.2305.08530}.

\bibitem[Arrow(1966)]{Arrow:1966}
Kenneth~J. Arrow.
\newblock Aspects of the theory of risk-bearing.
\newblock \emph{Economica}, 33:\penalty0 251, 1966.

\bibitem[Pratt(1964)]{Pratt:1964}
John~W. Pratt.
\newblock Risk aversion in the small and in the large.
\newblock \emph{Econometrica}, 32\penalty0 (1/2):\penalty0 122--136, 1964.

\bibitem[Black and Litterman(1990)]{BL:1990}
Fischer Black and Robert Litterman.
\newblock Asset allocation: Combining investor views with market equilibrium.
\newblock \emph{Goldman Sachs Fixed Income Research}, 1990.

\bibitem[Anderson(2003)]{Anderson:2003}
Theodore~W. Anderson.
\newblock An introduction to multivariate statistical analysis.
\newblock \emph{Wiley Interscience}, 2003.

\bibitem[Diamond and Boyd(2016)]{Diamond:2016}
Steven Diamond and Stephen Boyd.
\newblock {CVXPY}: {A} {P}ython-embedded modeling language for convex
  optimization.
\newblock \emph{Journal of Machine Learning Research}, 17\penalty0
  (83):\penalty0 1--5, 2016.

\bibitem[Ledoit and Wolf(2003)]{Ledoit:2003}
Olivier Ledoit and Michael Wolf.
\newblock Improved estimation of the covariance matrix of stock returns with an
  application to portfolio selection.
\newblock \emph{Journal of Empirical Finance}, 10\penalty0 (5):\penalty0
  603--621, 2003.

\bibitem[David M.~Corey(1998)]{Corey:1998}
Michael~Burke David M.~Corey, William P.~Dunlap.
\newblock Averaging correlations: Expected values and bias in combined pearson
  rs and fisher's z transformations.
\newblock \emph{The Journal of General Psychology}, 125\penalty0 (3):\penalty0
  245--261, 1998.

\bibitem[Archakov and Hansen(2022)]{Archakov:2022}
Ilya Archakov and Peter~R. Hansen.
\newblock A canonical representation of block matrices with applications to
  covariance and correlation matrices.
\newblock \emph{The Review of Economics and Statistics}, page 1–39, 2022.

\bibitem[LSE()]{LSE:wiki}
Wikipedia logsumexp.
\newblock URL \url{https://en.wikipedia.org/wiki/LogSumExp}.

\bibitem[Luxenberg and Boyd(2023)]{Boyd:2022}
Eric Luxenberg and Stephen Boyd.
\newblock Portfolio construction with gaussian mixture returns and exponential
  utility via convex optimization.
\newblock \emph{Optim Eng}, 2023.

\bibitem[Taraldsen(2023)]{Taraldsen:2023}
 Gunnar Taraldsen. 
 \newblock The Confidence Density for Correlation. 
 \newblock \emph{Sankhya} A \penalty0 85, \penalty0 600–616, 2023. 

\end{thebibliography}

\section*{Appendix A: Statistical uncertainty in variance and correlation coefficients}
In this section, we revisit well-established results concerning the finite sample uncertainty of variance and correlation coefficients, specifically in the idealized context of Gaussian returns. We anticipate that accounting for distribution with fatter tails would likely heighten the uncertainty estimates. Specifically, we focus on the conditional distribution of the population variance and correlation coefficient, given the sample variance and correlation coefficient. Although we do not apply the derived functional forms of uncertainty in the models discussed in this paper, we present them to illustrate how sample values may deviate from population parameters.

\subsection{Sample uncertainty of variance}

Assume that returns are i.i.d. and generated from a Gaussian distribution $r_t\sim N(\mu, \sigma^2)$. The joint distribution of $\mu$ and $\sigma^2$, given data $D=\{x_1,x_2,...,x_n\}$, is:
\be
p(\mu, \sigma^2|D)\sim \frac{1}{\sigma^{n+2}} e^{-\frac{\sum_{i=1}^{n} (x_i-\mu)^2}{2\sigma^2}} 
\ee

Assuming the mean is unknown and the prior for $\mu$ is $p(\mu)\sim \text{const}$, integrating over $\mu$ yields the marginal distribution of variance $\sigma^2$ given the sample variance estimate $s^2$:
\be 
p(\sigma^2|s^2)\sim (\sigma^2)^{-\frac{n+1}{2}} e^{-\frac{(n-1) s^2}{2 \sigma^2}}
\ee

This is equivalent to a scaled inverse chi-squared distribution $\chi^2(\nu,s^2)$ with parameters $\nu=n-1$ and scale $s^2=\frac{\sum_{i=1}^n (x_i-\bar{x})^2}{n-1}$.

\begin{figure}[t]
    \centering
    \includegraphics[width=\linewidth]{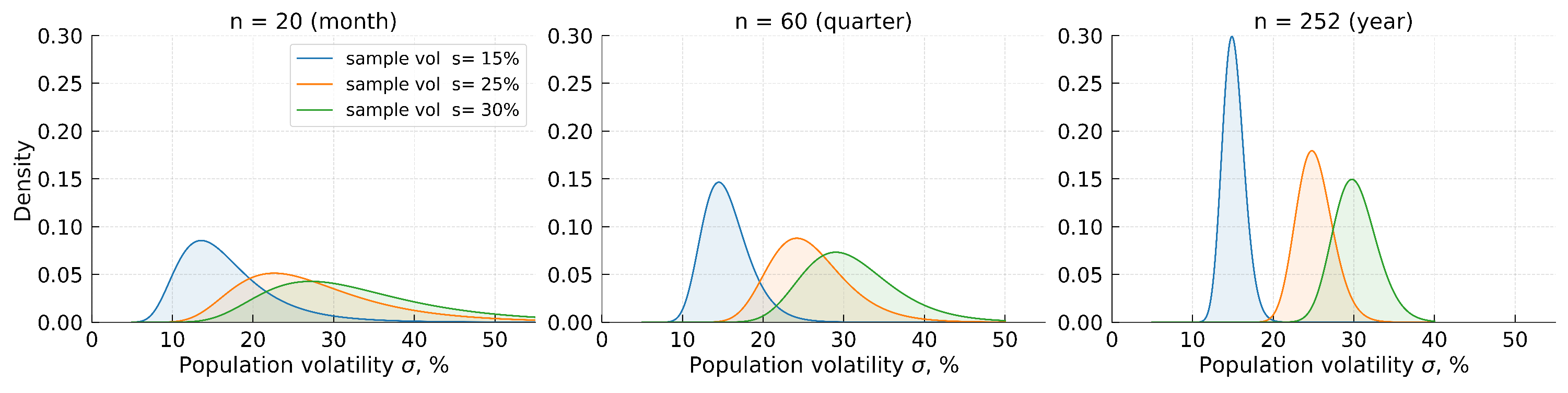}
    \caption{Conditional distribution of population volatility for a finite sample of observations}
    \label{fig:my_label_vol}
\end{figure}

In Figure~\ref{fig:my_label_vol}, for a commonly used values of $n=20$, $60$, $252$ days for volatility estimation, we present the distribution of the population volatility $\sigma$ corresponding to sample volatilities  $s=15\%$, $25\%$, and $30\%$. It can be observed that the variance distributions are broad and overlapping. Analytically, for $x\sim \chi^2(\nu,s^2)$, the mean is given by $E[x]=s^2 \frac{n-1}{n-3}$  and the variance by $Var[x]=s^4 \frac{2 (n-1)^2}{(n-3)^2 (n-5)}$. For $n=20$, the mean is  $\frac{19}{17}s^2$, and the standard deviation is $0.4 s^2$, with the ratio of the mean of the variance to its standard deviation being $\frac{19}{17}/0.4=2.8$.

\begin{figure}[t]
    \centering
    \includegraphics[width=\linewidth]{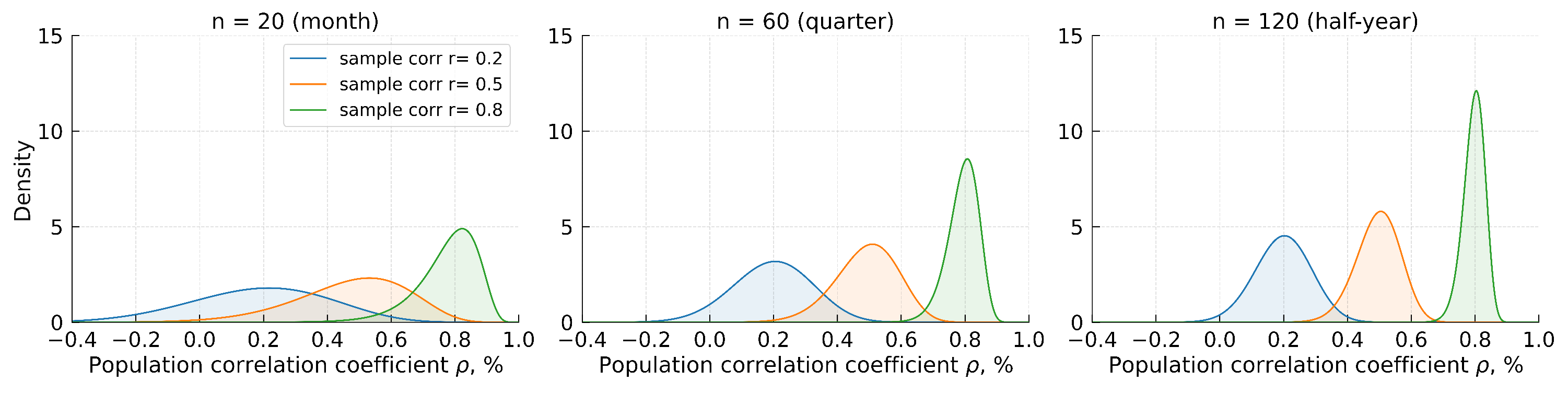}
    \caption{Conditional distribution of correlation coefficient for finite sample of observations}
    \label{fig:my_label_corr}
\end{figure}

\subsection{Sample uncertainty of correlation coefficient}

The conditional distribution of the correlation coefficient $\rho$, given the sample estimation $r$, for a bivariate normal distribution was calculated in~\cite{Taraldsen:2023}. The formula for this distribution is as follows:
\be
\pi(\rho|r)=\frac{\Gamma(\nu+1)}{\sqrt{2 \pi} \Gamma(\nu+\frac 12)} \left(1-r^2\right)^{\frac{\nu-1}{2}} \left(1-\rho^2\right)^{\frac{\nu-2}{2}}  \left(1-r \rho\right)^{\frac{1-2 \nu}{2}} {}_2F_1(\frac{3}{2},-\frac{1}{2},\nu+\frac{1}{2};\frac{1+r\rho}{2}) 
\ee
here, ${}_2F_1$ is the Gauss hypergeometric function and $\nu = n-1$. 

In Figure \ref{fig:my_label_corr}, we present the distribution of the population correlation coefficient $\rho$ for sample correlations of $r=0.2, 0.5$, and $0.8$ with sample sizes of $n = 20, 60$, and $120$.


In the financial industry, the convention is to use daily data with $n=20$ observations (a one-month window) for volatility estimation, and a year of data with $n=252$ observations for correlation and covariance matrix estimation. This approach can be seen as a practitioner's implicit trade-off between sample uncertainty and uncertainty caused by non-stationarity.

\section*{Appendix B: Distribution of parameters of two-dimensional Wishart matrices}

\begin{figure}[p]
    \centering
    \begin{minipage}{0.98\textwidth}
        \includegraphics[width=\linewidth]{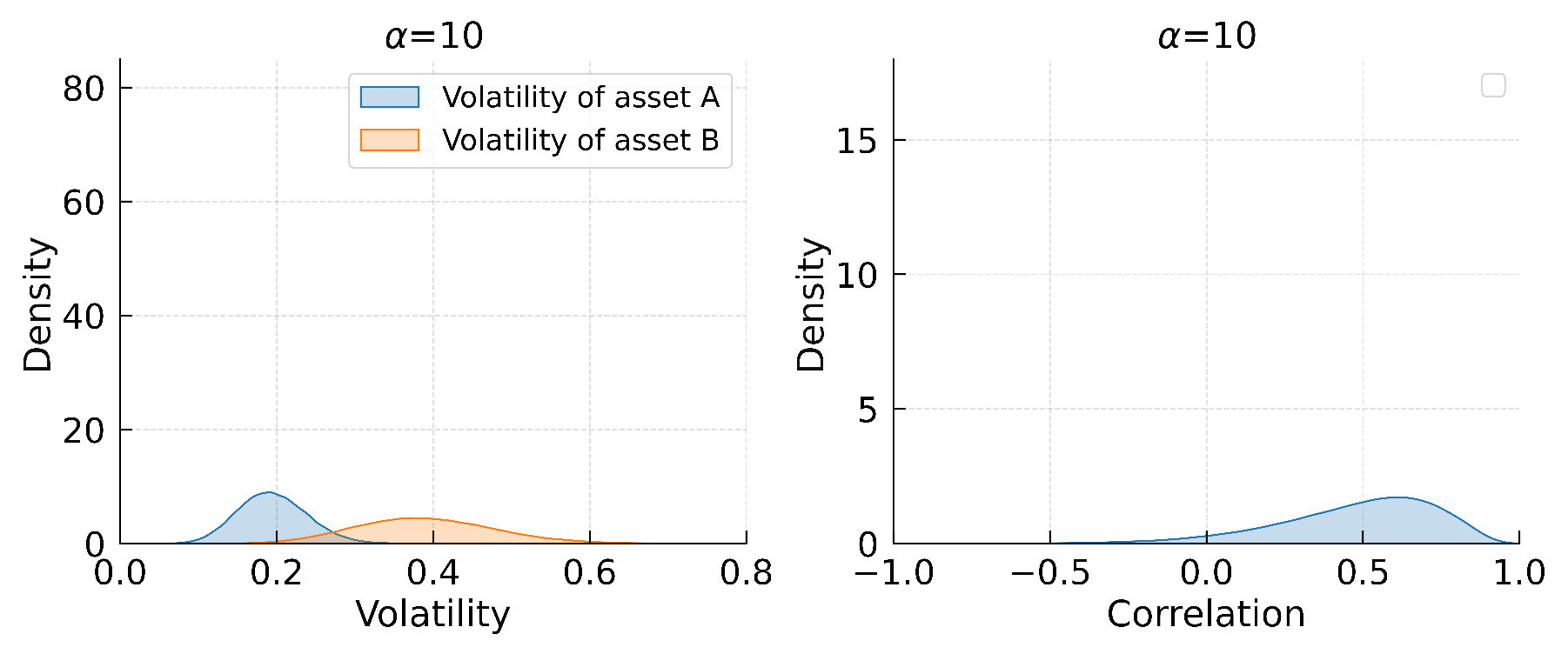}
    \end{minipage}
    
    
    \begin{minipage}{0.98\textwidth}
        \includegraphics[width=\linewidth]{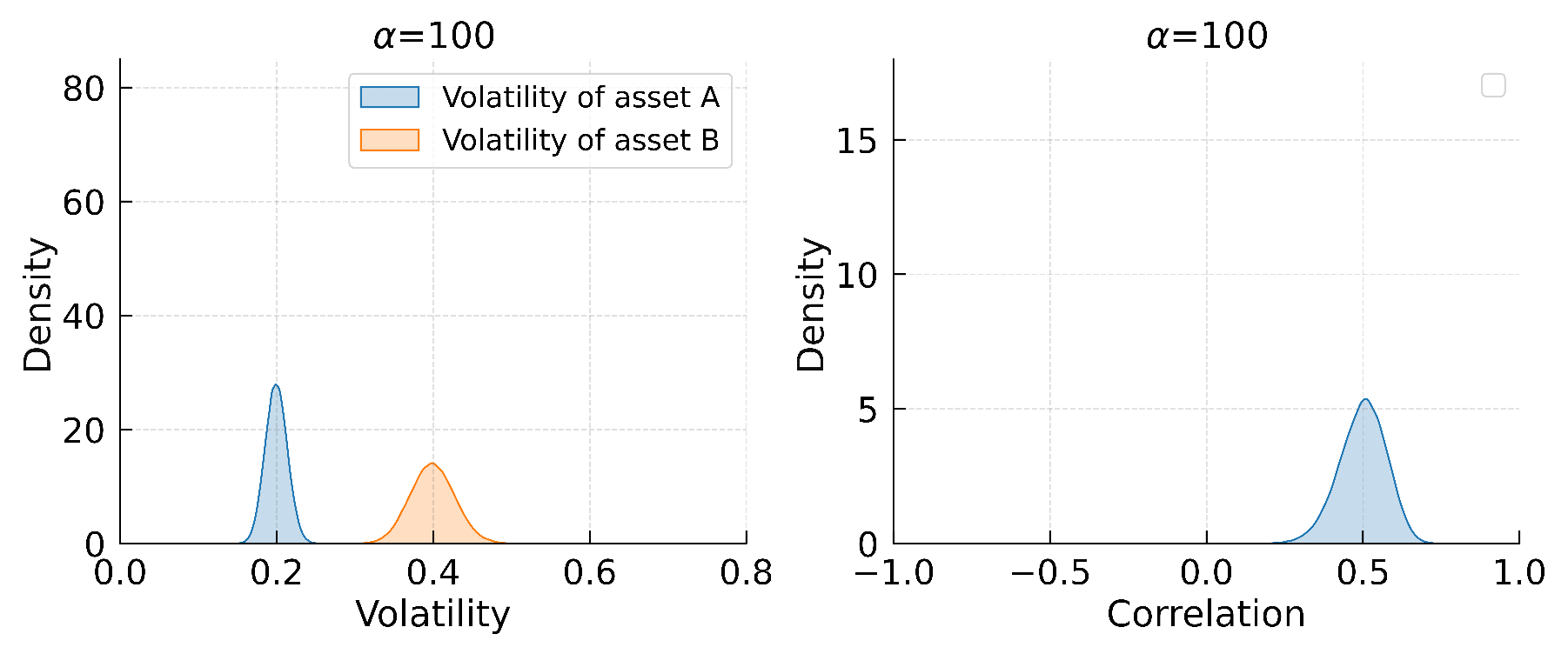}
    \end{minipage}

    
    \begin{minipage}{0.98\textwidth}
        \includegraphics[width=\linewidth]{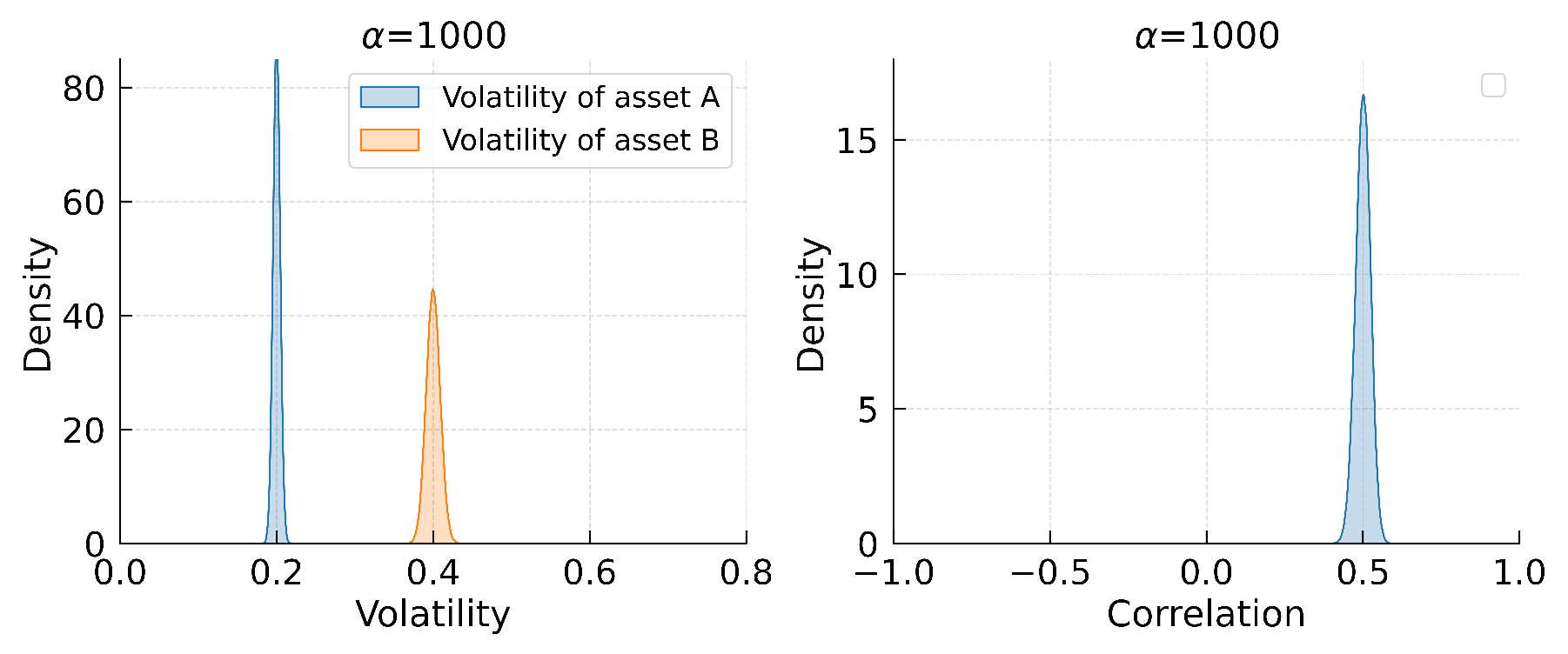}
    \end{minipage}
    
    \caption{Volatility (left panels) and correlation coefficient (right panels) distribution for $\alpha=10$ (high noise level, top panels), $\alpha=100$ (medium noise level, middle panels), and $\alpha=1000$ (low noise level, bottom panels)}
    \label{fig:three_images}
\end{figure}

In this section, we simulate the distribution of two-dimensional Wishart matrices as described by the model in Eq. \ref{eq:noise_m1}:
$$ 
\bs S \sim W_2(\alpha,\frac{\bs \Sigma}{\alpha})
\label{eq:noise_m1_appendix}
$$
with matrix $\bs \Sigma$  given by:
\[
\bs \Sigma =
\begin{bmatrix}
    \sigma_A^2 & \rho_{AB} \sigma_A \sigma_B \\
    \rho_{AB} \sigma_A \sigma_B  & \sigma_B^2 \\
\end{bmatrix}
\]
where the volatility of asset A is $\sigma_A=0.2$ (20\% annualized), that of asset B is $\sigma_B=0.4$ (40\% annualized), and the correlation coefficient is $\rho_{AB}=0.5$. We simulate $N=10^5$ samples and display the distribution of volatilities and the correlation coefficient. The result of the simulation is shown on Figure~\ref{fig:three_images}. 

\section{Appendix C: Dependence between volatility and correlation coefficient}

\begin{figure}[h]
    \centering
    \begin{subfigure}[b]{0.49\textwidth}
        \centering
        \includegraphics[width=1.35\textwidth]{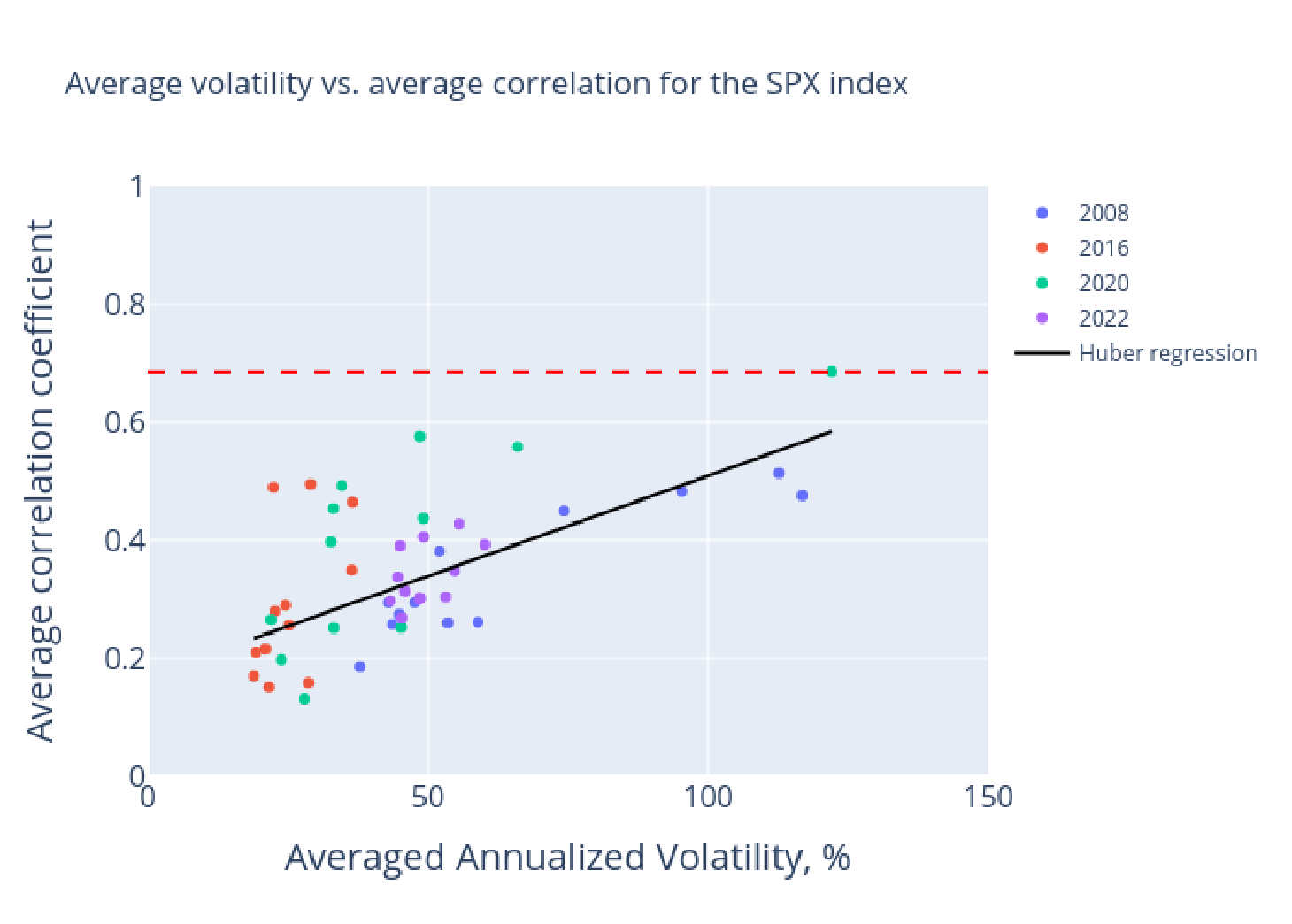}
    \end{subfigure}
    \hfill
    \begin{subfigure}[b]{0.49\textwidth}
        \centering
        \includegraphics[width=1.35\textwidth]{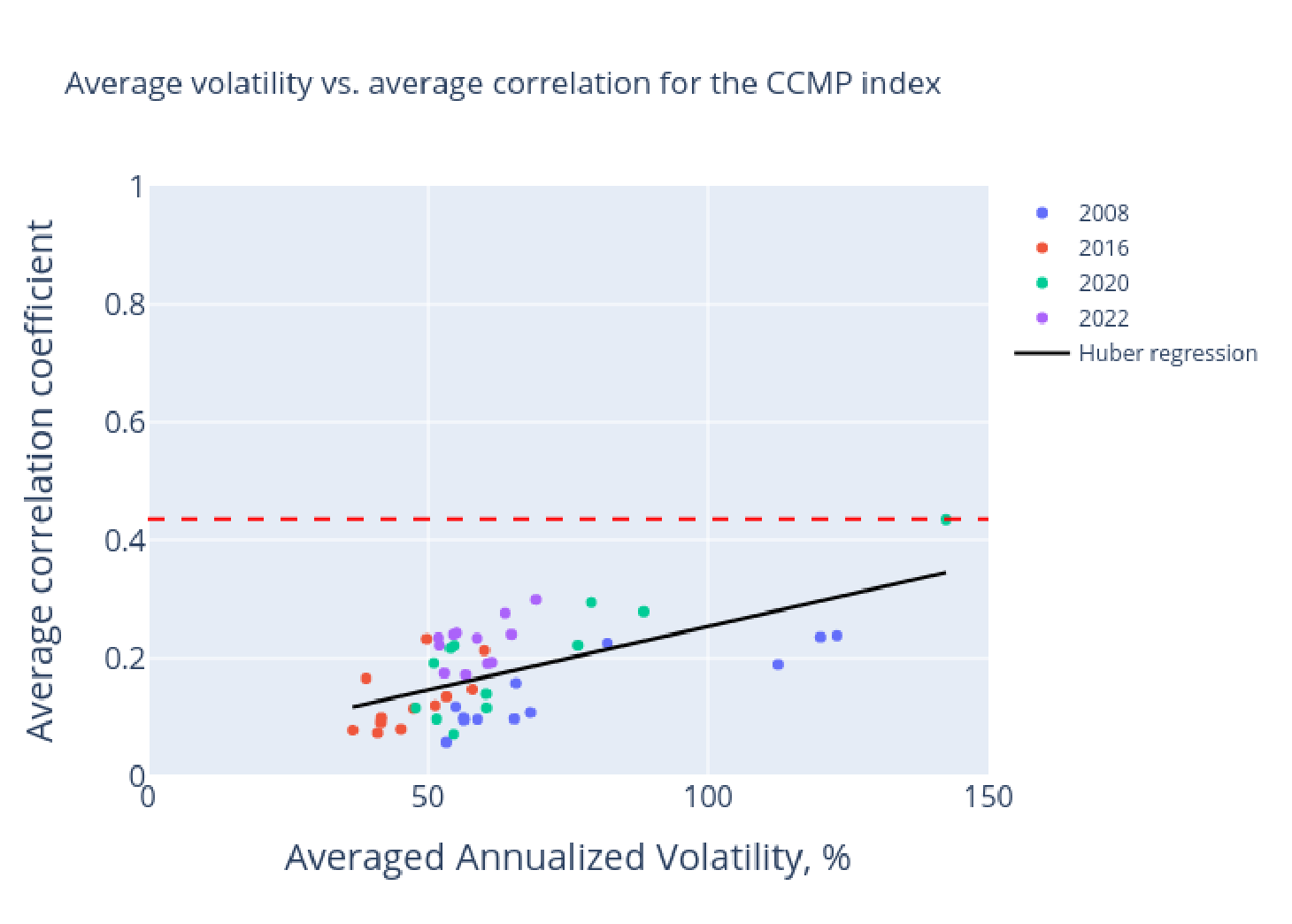}
    \end{subfigure}

\vspace{1cm}

    \begin{subfigure}[b]{0.49\textwidth}
        \centering
        \includegraphics[width=1.35\textwidth]{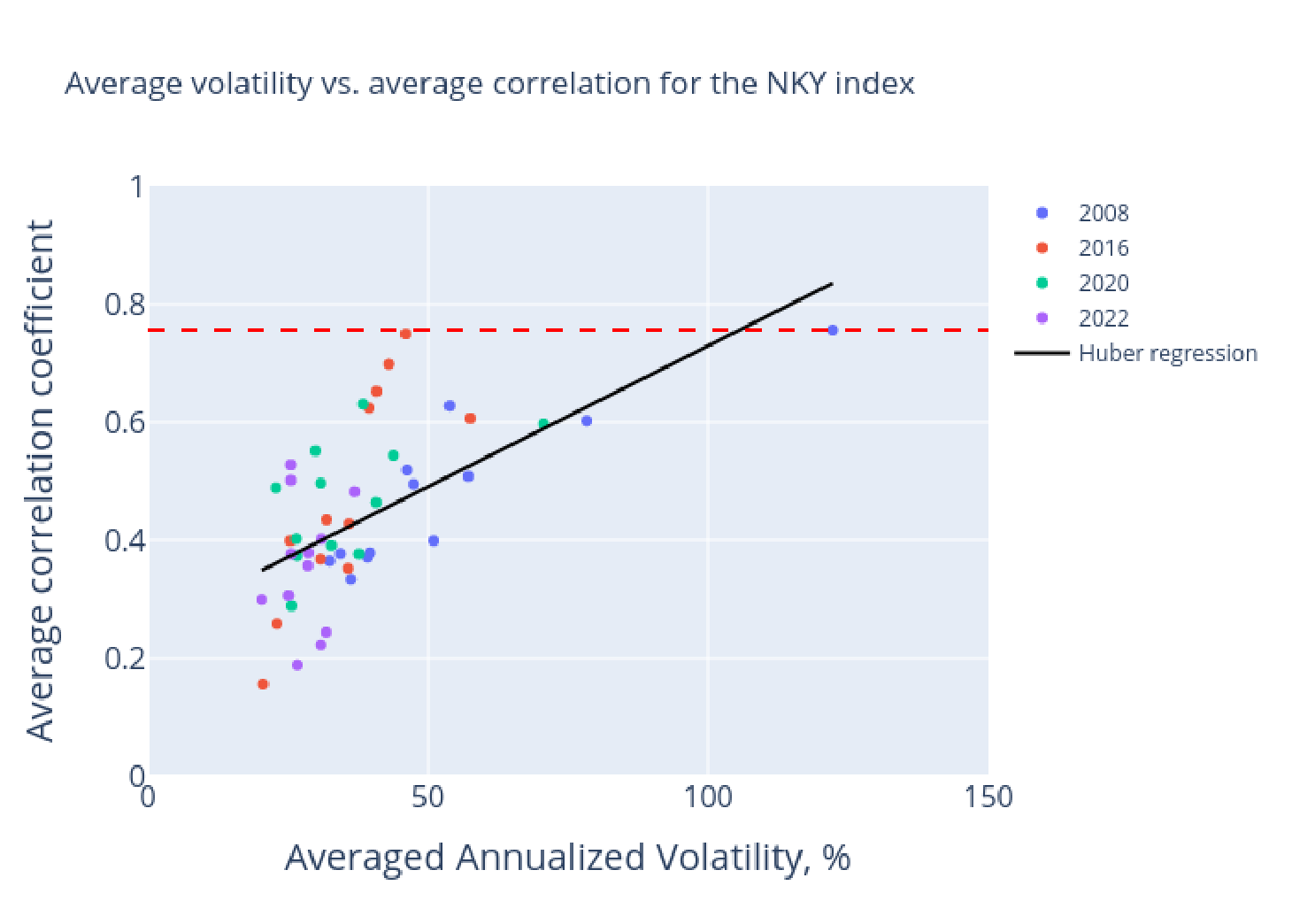}
    \end{subfigure}
    \hfill
    \begin{subfigure}[b]{0.49\textwidth}
        \centering
        \includegraphics[width=1.35\textwidth]{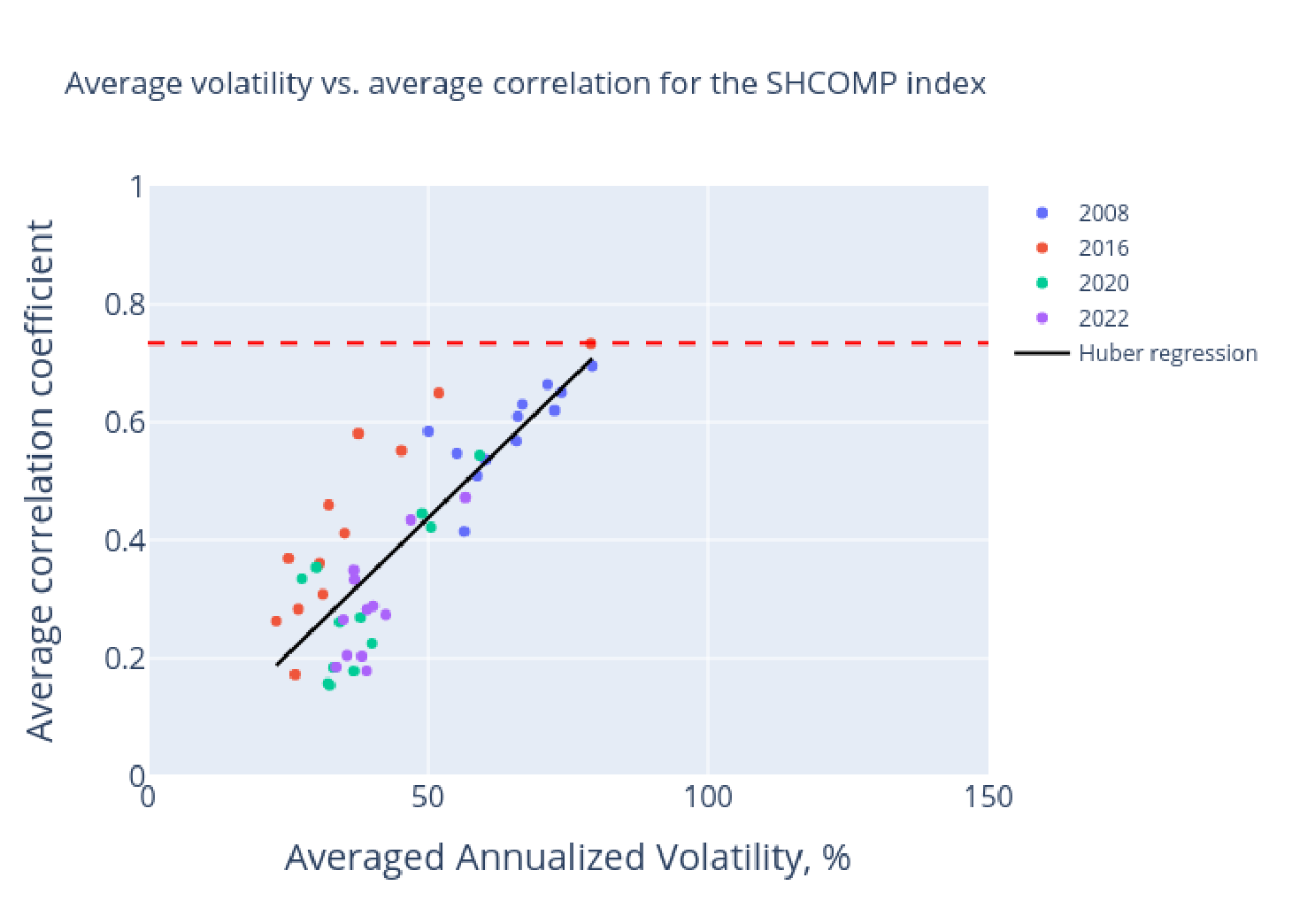}
    \end{subfigure}
    
    \caption{Scatter plot of average volatility versus average correlation coefficient for the SPX, CCMP, NKY, and SHCOMP indices. The red dashed line corresponds to the maximum correlation value, and the black solid line corresponds to the Huber regression curves.}
    \label{fig:vol_corr_subfigures}
\end{figure}

In this section, we calculated the average volatility and average correlation coefficient for major stock indexes on a monthly scale for the years 2008, 2016, 2020, and 2022. We then plotted these values on a scatter plot Figure \ref{fig:vol_corr_subfigures}. Figures for all indexes are shown in \footnote{Figures for all indices: \href{https://github.com/maxmarkov/portfolio-uncertain-covariance/blob/main/notebook-volatility-vs-correlation-results.ipynb}{link to the Supplementary Material page on GitHub.}}. The constituents of the index were taken from the first day of each respective year. This result can be used to model the covariance matrix in the stressed regime $\bs \Sigma_s$ in equivariant and equicorrelation approximation.  

\end{document}